\documentclass[mathleft
]{an}
\usepackage{epsfig}
\usepackage{epstopdf}
\usepackage{graphicx}
\usepackage{times}
\usepackage{natbib}
\bibpunct{(}{)}{;}{a}{}{,}
\begin{document}


\title{Might intermediate-order g modes observed in the \emph{CoRoT} hybrid \\
$\gamma$ Doradus/$\delta$ Scuti star HD 49434 be stochastically excited?}

\author{
T. L. Campante\inst{1,2}\fnmsep\thanks{\email{campante@astro.up.pt}\newline}
\and A. Grigahc\`ene\inst{1}
\and J. C. Su\'arez\inst{3}
\and M. J. P. F. G. Monteiro\inst{1,4}
}
\titlerunning{Might intermediate-order g modes in the hybrid HD 49434 be stochastically excited?}
\authorrunning{T. L. Campante et al.}
\institute{
Centro de Astrof\'{\i}sica da Universidade do Porto, Rua das Estrelas, 4150-762 Porto, Portugal
\and 
Danish AsteroSeismology Centre, Aarhus University, Ny Munkegade 120, DK-8000 Aarhus C, Denmark
\and 
Instituto de Astrof\'isica de Andaluc\'ia (CSIC), Apartado 3004, 18080 Granada, Spain
\and
Departamento de F\'{\i}sica e Astronomia, Faculdade de Ci\^encias, Universidade do Porto, Portugal
}

\received{}
\accepted{}
\publonline{}


\abstract{
The primary target in the seismo-field of \emph{CoRoT}, HD 49434, known to be a bona fide hybrid $\gamma$ Doradus/$\delta$ Scuti, shows excited intermediate-order g modes. Time-Dependent Convection models, however, predict a range in frequency that is stable to pulsations, between the simultaneously excited high-order g modes ($\gamma$ Dor) and low-order p and g modes ($\delta$ Sct). Furthermore, theoretical studies based on model computations of $\delta$ Sct stars suggest that stochastically excited modes are likely to be observed. A pertinent question would then be to ask: Might those observed intermediate-order g modes be stochastically excited? By employing a statistical method which searches for the signature of stochastic excitation in stellar pulsations, we investigate the nature of those modes with possible implications on the identification of their excitation mechanism. Preliminary results are rather inconclusive about the presence of stochastic excitation. A new analysis that thoroughly takes into account sampling effects is necessary in order to get more reliable results.
}

\maketitle

\section{Introduction}

Hybrid $\gamma$ Dor/$\delta$ Sct stars are of great interest because they offer additional constraints on stellar structure and may be used to test theoretical models. $\gamma$ Dor stars pulsate in high-order g modes with periods of the order of 1 day, driven by convective flux blocking at the base of their convective envelopes. In turn, $\delta$ Sct stars pulsate in low-order p and g modes with periods of the order of 2 hours, driven by the $\kappa$-mechanism operating in the \ion{He}{ii} ionization zone. Most of the already confirmed hybrid stars clearly display separate frequency domains. However, some of these objects are known to exhibit a range of excited intermediate modes. Such a feature has been detected for the first time in the frequency power spectrum of the Am star HD 8801 by \citet{Handler}. That work also shows that the amplitudes of these intermediate oscillation modes are of the same order of magnitude as of those detected in the classical $\delta$ Sct domain. The presence of intermediate-order modes opens new questions concerning the possible mechanism responsible for their excitation.

In this regard, the present work aims at providing insight into the possible excitation mechanism of such intermediate-order modes, through the analysis of the frequency power spectrum of the hybrid star HD 49434, which has been selected for the asteroseismic core programme of the \emph{CoRoT} satellite \citep{CoRoT}. HD 49434 (spectral type F1V) is a bright ($V\!=\!5.75$) and multiperiodic pulsator located near the blue edge of the $\gamma$ Dor instability strip (IS) and inside the $\delta$ Sct IS. Photometric data recently made available by \emph{CoRoT} \citep{Chapellier} allowed confirming the hybrid nature of HD 49434. This star had been referenced as a candidate hybrid $\gamma$ Dor/$\delta$ Sct by \citet{Uytterhoeven}, following an extensive photometric and spectroscopic ground-based campaign. A compelling feature of its frequency power spectrum is the presence of excited intermediate-order g modes.

From a theoretical point of view, different approaches are currently being adopted in order to explain such spectral features. Theoretical studies based on model computations of cool $\delta$ Sct stars \citep{Houdek,Samadi} predict the presence of solar-like oscillations with amplitudes large enough to be detectable with ground-based instruments, although so far not confirmed. It should be stressed here that these models lie exactly on or near the $\gamma$ Dor IS determined afterwards by \citet{Dupret}. Therefore, it is plausible considering that solar-like, $\gamma$ Dor-like and $\delta$ Sct-like oscillations might be simultaneously excited in such intermediate-mass, main-sequence stars. 

Rotational splitting has already been invoked as a possible explanation for these observed frequencies \citep{Uytterhoeven,Bouabid}. More recently, \citet{Kallinger} suggested that most of the peaks in the rich frequency spectra of the $\delta$ Sct stars already analysed using data from \emph{CoRoT} \citep[see e.g.,][]{GH} could be the signature of non-white granulation background noise, meaning that only a few tens of those frequencies should be interpreted as stellar p modes.

On the other hand, based on the Time-Dependent Convection (TDC) of \citet{Grig05}, \citet{Dupret} showed that TDC models predict the likely existence of hybrid stars with both $\delta$ Sct p-mode and $\gamma$ Dor g-mode oscillations. Furthermore, TDC models predict the existence of a frequency \emph{gap} that is stable to pulsations in the range 5--15$\:\rm{d^{-1}}$ for low-degree modes. Therefore, according to these models, neither the classical $\kappa$-mechanism nor the convective flux blocking at the bottom of the convective envelope can explain the excitation of observed intermediate-order g modes.  

In the present work, we address the possibility that such intermediate-order g modes are excited by a stochastic mechanism. In order to do so, a first approach to the problem can be easily implemented by considering the characteristics of this type of excitation, in particular those concerning the statistical behaviour of the mode amplitudes. A simple diagnostic method has been established by \citet{Pereira05} that probes the stellar pulsations' excitation mechanism by analysing the temporal variation in the amplitude of oscillation modes. Numerical simulations and the application to the $\gamma$ Dor star HD 22702 \citep{Pereira07} serve as a test of the method. In this work we employ this statistical method to investigate the nature of the intermediate-order g modes visible in the spectrum of HD 49434.

\section{Observations vs.~TDC models}
\begin{figure}[h]
\centering
\resizebox{\hsize}{!}{\includegraphics{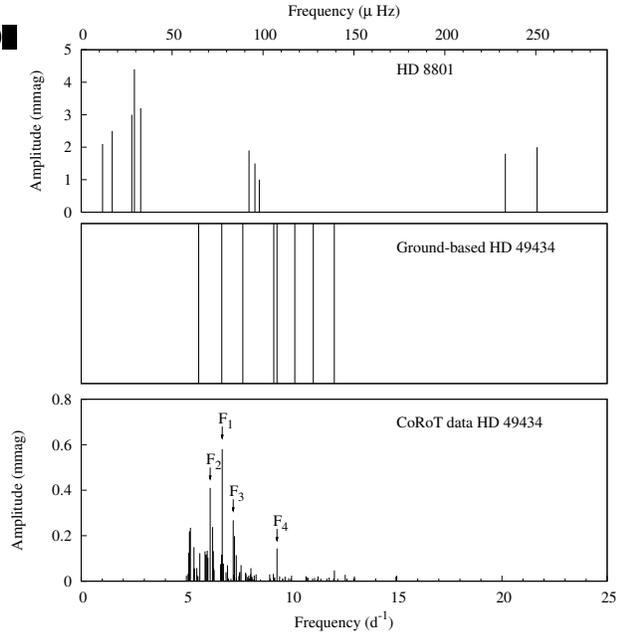}} 
\caption{
\emph{Top Panel:} Frequency spectrum of HD 8801 where $v$-filter amplitudes are given \citep{Handler}. \emph{Middle Panel:} Vertical lines represent ground-based observed frequecies for HD 49434 \citep{Uytterhoeven}. \emph{Bottom Panel:} Partial \emph{CoRoT} frequency spectrum of HD 49434. The 4 oscillation modes selected in \S\ref{sec:test} for the purpose of conducting the statistical test are indicated.
}
\label{spec1}
\end{figure}

Recent data from \emph{CoRoT} allowed confirming the hybrid nature of HD 49434, previously referenced as a candidate. In Fig.~\ref{spec1} we simultaneously display partial (i.e., exclusively within the \emph{gap}) ground-based and \emph{CoRoT} frequency spectra of HD 49434 (middle and bottom panels, respectively). The top panel displays the frequency spectrum of HD 8801.

\begin{figure}[h]
\centering
\resizebox{\hsize}{!}{\includegraphics{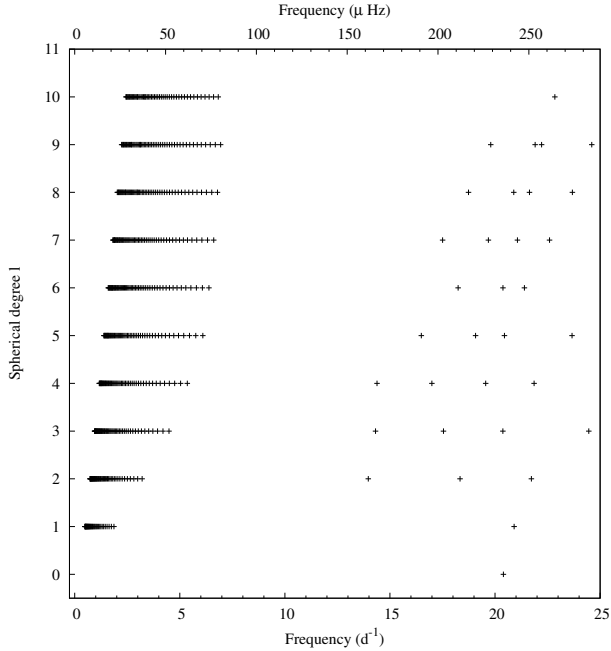}} 
\caption{Theoretical excited frequency spectrum for spherical degree $\ell \! \in \! [0,10]$. The theoretical stable frequency \emph{gap} is clearly seen even for high-degree modes.
}
\label{spec2}
\end{figure}

Making use of the fundamental parameters of HD 49434 ($T_{\mathrm{eff}}\!=\!7300 \pm 200\:\mathrm{K}$, $\log g\!=\!4.1 \pm 0.2$, $[\mathrm{Fe}/\mathrm{H}]\!=\!-0.1 \pm 0.2$, and $v\sin i\!=\!84 \pm 5\:\mathrm{km\,s^{-1}}$) taken from \citet{Uytterhoeven}, we are able to show in Fig.~\ref{spec2} the theoretical excited frequency spectrum for spherical degree $\ell$ ranging from 0 to 10 and corresponding to a representative model ($M\!=\!1.55\:\mathrm{M}_\odot$ and $T_{\mathrm{eff}}\!=\!7300\:\mathrm{K}$). We note that the stability of modes within the \emph{gap} decreases as degree $\ell$ increases. This is due to the fact that the Lamb frequency ($S_\ell$) behaves according to $S^{2}_{\ell} \propto \ell (\ell+1)$, so that the size of the propagation cavity for g modes increases with $\ell$ for a given frequency \citep{Grig10}. The \emph{gap} extent gets clearly reduced for higher values of $\ell$ without however becoming completely filled.

This interestingly opens a new question both on the validity of the data analysis treatment used in frequency detection and on the nature of the excitation mechanism once the stellar origin of these frequencies is assured.

\section{Search for evidence of stochastic excitation}\label{sec:test}

A search for the signature of stochastic excitation in a number of modes within the \emph{gap} was carried out according to the statistical method described in \citet{Pereira05}. It has been shown that for oscillations that are excited and damped by a physical process in stochastic equilibrium, the ratio of the standard deviation of the amplitude, $\sigma(A)$, over the amplitude mean value, $\langle A \rangle$, is approximately 0.52. This theoretical relation still holds true in the presence of a time series crowded with closely-spaced periods, a relevant issue when applying the method to $\gamma$ Dor stars. Furthermore, the region for which $\sigma(A) < 0.52 \langle A \rangle$ corresponds to oscillations excited by thermal overstability, this being the region where one expects to find most of the modes excited by the $\kappa$-mechanism. On the other hand, the region with $\sigma(A) > 0.52 \langle A \rangle$ corresponds to non-equilibrium stochastic oscillations (yet to be observed).

In order to investigate the nature of the observed intermediate-order g modes we have analysed the 136.9-day time series available for this star which was collected by \emph{CoRoT} during the long run LRa01 (2007 October--2008 March). We started by dividing the original time series in 41 subseries, each with the approximate duration of $3.33\:\mathrm{d}$. We then selected 4 previously detected modes within the \emph{gap}, having performed 41 amplitude measurements for each, i.e., one measurement per subseries. This allowed us to compute the statistic $\sigma(A)/\langle A \rangle$ (see Table \ref{tab2}).

\begin{figure}[h]
     \centering
 \resizebox{\hsize}{!}{\includegraphics{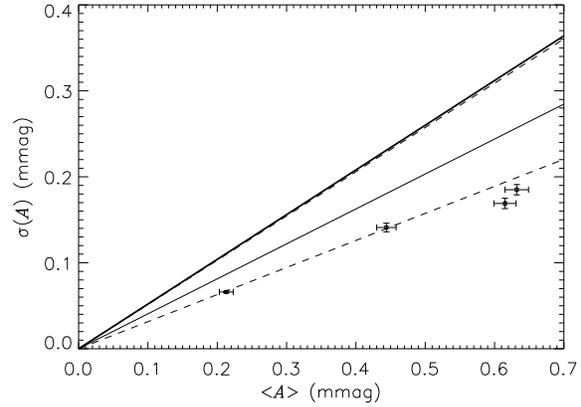}} 
      \caption{Excitation diagram: The thick solid line represents the theoretical relation $\sigma(A)=0.52\langle A \rangle$, whereas the thin solid line represents the outcome of the simulations that gave $\sigma(A) \approx 0.41\langle A \rangle$. The dashed lines represent the $1\sigma$ bounds for $\sigma(A)/\langle A \rangle$ taken from the pdf. Observational results for the selected modes are plotted with accompanying error bars.}
     \label{excitation}
\end{figure}

We went further and performed Monte Carlo simulations in order to compute the probability density function (pdf) for the observed $\sigma(A)/\langle A \rangle$ statistic assuming stochastic excitation. The modal bin of this distribution turns out to be approximately 0.407 and the $1\sigma$-equivalent confidence interval is given by $0.315 \leq \sigma(A)/\langle A \rangle \leq 0.515$. Fig.~\ref{excitation} displays the so-called excitation diagram. 

\begin{table}[h]  
\centering      
\caption{The statistic $\sigma(A)/\langle A \rangle$ computed along the 41 subseries for 4 selected modes within the \emph{gap}. The software package {\sc{Period04}} \citep{Lenz} was used to perform the amplitude measurements.}
\begin{tabular}{c c c c c c}  
\hline\hline                        
& Frequency & $\sigma(A)$ & $\langle A \rangle$ & $\sigma(A)/\langle A \rangle$ \\   
& ($\mathrm{d^{-1}}$) & (mmag) & (mmag) & \\ [1ex] 
\hline                   
$F_{1}$ & 6.703 & 0.185 & 0.632 & $0.293 \pm 0.013$ \\ [0.5ex]   
$F_{2}$ & 6.130 & 0.169 & 0.615 & $0.275 \pm 0.013$ \\ [0.5ex]
$F_{3}$ & 7.226 & 0.141 & 0.444 & $0.318 \pm 0.015$ \\ [0.5ex]
$F_{4}$ & 9.307 & 0.066 & 0.213 & $0.311 \pm 0.016$ \\ [1ex]
\hline     
\end{tabular}  
\label{tab2}
\end{table} 

\section{Discussion and preliminary conclusions}

Two remarks should be made after inspection of the excitation diagram (Fig.~\ref{excitation}):
\begin{enumerate}
\item The way the sampling has been performed, i.e., the number of amplitude measurements and the type of sampling (linear with separation 1, in other words, sampling from contiguous subseries), does not allow for a good estimate of $\sigma(A)/\langle A \rangle$ if we assume stochastic excitation. 
\item The fact that the observational results lie outside the confidence interval (for the two strongest modes) or just on the lower $1\sigma$ bound (for the two faintest modes) might tempt us to conclude that these modes are not stochastically excited. However, we need to be cautious and a new analysis should definitely be carried out where we increase the number of amplitude measurements (with the corresponding loss in frequency resolution of each of the subseries), thus increasing the significance of the statistic $\sigma(A)/\langle A \rangle$.
\end{enumerate}

\acknowledgements
This work was supported by the European Helio- and Asteroseismology Network (HELAS), a major international collaboration funded by the European Commission's Sixth Framework Programme. TLC is supported by grant with reference number SFRH/BD/36240/2007 from FCT/MCTES, Portugal. AG is supported by grant with reference number SFRH/BPD/41270/2007 from FCT/MCTES, Portugal. AG and MJPFGM are co-supported by project PTDC/CTE-AST/098754/2008 from FCT/MCTES, Portugal. JCS acknowledges support from the Instituto de Astrof\'{\i}sica de Andaluc\'{\i}a (CSIC) through an Excellence Project postdoctoral fellowship financed by the Spanish Consejer\'{\i}a de Innovaci\'on, Ciencia y Empresa de la Junta de Andaluc\'{\i}a with reference number FQM4156-2008.


\bibliographystyle{astron}
\bibliography{bibliography}

\begin{thebibliography}{}

\bibitem[\protect\astroncite{{Baglin} et~al.}{2006}]{CoRoT}
{Baglin}, A., {Michel}, E., {Auvergne}, M., and {The COROT Team}: 2006,
\newblock in {\em Proceedings of SOHO 18/GONG 2006/HELAS I, Beyond the
  spherical Sun}, Vol. 624 of {\em ESA Special Publication}

\bibitem[\protect\astroncite{{Bouabid} et~al.}{2009}]{Bouabid}
{Bouabid}, M., {Montalb{\'a}n}, J., {Miglio}, A., {Dupret}, M.,
  {Grigahc{\`e}ne}, A., and {Noels}, A.: 2009,
\newblock in {J.~A.~Guzik \& P.~A.~Bradley} (ed.), {\em American Institute of
  Physics Conference Series}, Vol. 1170 of {\em American Institute of Physics
  Conference Series}, pp 477--479

\bibitem[\protect\astroncite{{Chapellier} et~al.}{2009}]{Chapellier}
{Chapellier}, E., {Bouabid}, M., {Le Contel}, D., {Rodriguez}, E., and
  {Mathias}, P.: 2009,
\newblock in {J.~A.~Guzik \& P.~A.~Bradley} (ed.), {\em American Institute of
  Physics Conference Series}, Vol. 1170 of {\em American Institute of Physics
  Conference Series}, pp 472--473

\bibitem[\protect\astroncite{{Dupret} et~al.}{2005}]{Dupret}
{Dupret}, M., {Grigahc{\`e}ne}, A., {Garrido}, R., {Gabriel}, M., and
  {Scuflaire}, R.: 2005,
\newblock {\em A\&A} {\bf 435}, 927

\bibitem[\protect\astroncite{{Garc{\'{\i}}a Hern{\'a}ndez} et~al.}{2009}]{GH}
{Garc{\'{\i}}a Hern{\'a}ndez}, A., {Moya}, A., {Michel}, E., {Garrido}, R.,
  {Su{\'a}rez}, J.~C., {Rodr{\'{\i}}guez}, E., {Amado}, P.~J.,
  {Mart{\'{\i}}n-Ruiz}, S., {Rolland}, A., {Poretti}, E., {Samadi}, R.,
  {Baglin}, A., {Auvergne}, M., {Catala}, C., {Lefevre}, L., and {Baudin}, F.:
  2009,
\newblock {\em A\&A} {\bf 506}, 79

\bibitem[\protect\astroncite{{Grigahc{\`e}ne} et~al.}{2010}]{Grig10}
{Grigahc{\`e}ne}, A., {Antoci}, V., {Balona}, L., {Catanzaro}, G.,
  {Daszynska-Daszkiewicz}, J., {Guzik}, J.~A., {Handler}, G., {Houdek}, G.,
  {Kurtz}, D.~W., {Marconi}, M., {Monteiro}, M.~J.~P.~F.~G., {Moya}, A.,
  {Ripepi}, V., {Suarez}, J., {Uytterhoeven}, K., {Borucki}, W.~J., {Brown},
  T.~M., {Christensen-Dalsgaard}, J., {Gilliland}, R.~L., {Jenkins}, J.~M.,
  {Kjeldsen}, H., {Koch}, D., {Bernabei}, S., {Bradley}, P., {Breger}, M., {Di
  Criscienzo}, M., {Dupret}, M., {Garcia}, R.~A., {Garcia Hernandez}, A.,
  {Jackiewicz}, J., {Kaiser}, A., {Lehmann}, H., {Marin-Ruiz}, S., {Mathias},
  P., {Molenda-Zakowicz}, J., {Nemec}, J.~M., {Nuspl}, J., {Paparo}, M.,
  {Roth}, M., {Szabo}, R., {Suran}, M.~D., and {Ventura}, R.: 2010,
\newblock {\em ArXiv e-prints}

\bibitem[\protect\astroncite{{Grigahc{\`e}ne} et~al.}{2005}]{Grig05}
{Grigahc{\`e}ne}, A., {Dupret}, M., {Gabriel}, M., {Garrido}, R., and
  {Scuflaire}, R.: 2005,
\newblock {\em A\&A} {\bf 434}, 1055

\bibitem[\protect\astroncite{{Handler}}{2009}]{Handler}
{Handler}, G.: 2009,
\newblock {\em MNRAS} {\bf 398}, 1339

\bibitem[\protect\astroncite{{Houdek} et~al.}{1999}]{Houdek}
{Houdek}, G., {Balmforth}, N.~J., {Christensen-Dalsgaard}, J., and {Gough},
  D.~O.: 1999,
\newblock {\em A\&A} {\bf 351}, 582

\bibitem[\protect\astroncite{{Kallinger} and {Matthews}}{2010}]{Kallinger}
{Kallinger}, T. and {Matthews}, J.~M.: 2010,
\newblock {\em ApJ} {\bf 711}, L35

\bibitem[\protect\astroncite{{Lenz} and {Breger}}{2005}]{Lenz}
{Lenz}, P. and {Breger}, M.: 2005,
\newblock {\em Communications in Asteroseismology} {\bf 146}, 53

\bibitem[\protect\astroncite{{Pereira} and {Lopes}}{2005}]{Pereira05}
{Pereira}, T.~M.~D. and {Lopes}, I.~P.: 2005,
\newblock {\em ApJ} {\bf 622}, 1068

\bibitem[\protect\astroncite{{Pereira} et~al.}{2007}]{Pereira07}
{Pereira}, T.~M.~D., {Su{\'a}rez}, J.~C., {Lopes}, I., {Mart{\'{\i}}n-Ruiz},
  S., {Amado}, P.~J., {Garrido}, R., {Rodr{\'{\i}}guez}, E., {Costa}, V.,
  {Rolland}, A., {Arellano Ferro}, A., and {Sareyan}, J.: 2007,
\newblock {\em A\&A} {\bf 464}, 659

\bibitem[\protect\astroncite{{Samadi} et~al.}{2002}]{Samadi}
{Samadi}, R., {Goupil}, M., and {Houdek}, G.: 2002,
\newblock {\em A\&A} {\bf 395}, 563

\bibitem[\protect\astroncite{{Uytterhoeven} et~al.}{2008}]{Uytterhoeven}
{Uytterhoeven}, K., {Mathias}, P., {Poretti}, E., {Rainer}, M.,
  {Mart{\'{\i}}n-Ruiz}, S., {Rodr{\'{\i}}guez}, E., {Amado}, P.~J., {Le
  Contel}, D., {Jankov}, S., {Niemczura}, E., {Pollard}, K.~R., {Brunsden}, E.,
  {Papar{\'o}}, M., {Costa}, V., {Valtier}, J., {Garrido}, R., {Su{\'a}rez},
  J.~C., {Kilmartin}, P.~M., {Chapellier}, E., {Rodr{\'{\i}}guez-L{\'o}pez},
  C., {Marin}, A.~J., {Aceituno}, F.~J., {Casanova}, V., {Rolland}, A., and
  {Olivares}, I.: 2008,
\newblock {\em A\&A} {\bf 489}, 1213

\end{thebibliography}

\end{document}